\documentclass{article}
\usepackage{graphicx}
\usepackage{slashed}
\usepackage{amsmath}

\def\beq{\begin{eqnarray}}
\def\eeq{\end{eqnarray}}
\def\bea{\begin{eqnarray*}}
\def\eea{\end{eqnarray*}}

\def\centeron#1#2{{\setbox0=\hbox{#1}\setbox1=\hbox{#2}\ifdim
\wd1>\wd0\kern.5\wd1\kern-.5\wd0\fi
\copy0\kern-.5\wd0\kern-.5\wd1\copy1\ifdim\wd0>\wd1
\kern.5\wd0\kern-.5\wd1\fi}}
\def\ltap{\;\centeron{\raise.35ex\hbox{$<$}}{\lower.65ex\hbox{$\sim$}}\;}
\def\gtap{\;\centeron{\raise.35ex\hbox{$>$}}{\lower.65ex\hbox{$\sim$}}\;}

\def\singleandthirdspaced{\baselineskip=\normalbaselineskip\multiply
    \baselineskip by 130\divide\baselineskip by 100}

\newcommand{\newc}{\newcommand}
\newc{\qbar}{{\overline q}}
\newc{\Kahler}{K\"ahler }
\newc{\deltaGS}{\delta_{\rm GS}}

\newcommand{\NN}{\mathcal{N}}
\newcommand{\PP}{\mathcal{P}}
\newcommand{\RR}{\mathcal{R}}

\begin{document}
\begin{titlepage}
\begin{flushright}
{\large 
SCIPP 13/15\\
}
\end{flushright}

\vskip 1.2cm

\begin{center}

{\LARGE\bf Small Field Inflation and the Spectral Index}

\vskip 1.4cm

{\large  Milton Bose, Michael Dine, Angelo Monteux and Laurel Stephenson Haskins}
\\
\vskip 0.4cm
{\it Santa Cruz Institute for Particle Physics and
\\ Department of Physics,
     Santa Cruz CA 95064  } \\
\vskip 4pt

\vskip 1.5cm

\begin{abstract}
It is sometimes stated that $n_s = 0.98$ in hybrid inflation; sometimes that it predicts $n_s >1$.  A number of authors have consider
aspects of Planck scale corrections and argued that they affect these predictions.  Here we consider these systematically, describing
the situations which can yield $n_s =0.96$, and the extent to which this result requires additional tuning.  \end{abstract}

\end{center}

\vskip 1.0 cm

\end{titlepage}
\setcounter{footnote}{0} \setcounter{page}{2}
\setcounter{section}{0} \setcounter{subsection}{0}
\setcounter{subsubsection}{0}

\singleandthirdspaced

\section{Introduction}
\label{introduction}

In \cite{dinepack}, it was argued that,
 with some very mild assumptions about genericity, we can characterize small field inflation quite simply.  First, it was argued that the 
 effective theory should exhibit an approximate (global) supersymmetry in order that there be fields light on the scale of the Hubble
 constant during inflation, $H_I$.  Then, assuming $H_I \gg m_{3/2}$:
\begin{enumerate}
\item
The inflaton is a pseudomodulus, labeling a set of approximate ground states with spontaneously broken
supersymmetry.
\item
The effective theory should obey a discrete $R$ symmetry in order that the cosmological constant (c.c.) be approximately
zero at the end of inflation.
\item
At the end of inflation, the inflaton must couple through relevant or marginal operators to fields which are light with respect to the scale of the energy density during inflation, in order that the cosmological constant be small at the end of inflation.  In particular, it was stressed that
inflation typically ends, in the hybrid case, {\it before the inflaton reaches the waterfall region}.
\end{enumerate}
So-called models of hybrid inflation\cite{linderiotto,lindehybrid,hybrid1,hybrid2,hybrid3} have in common the last feature above; in \cite{dinepack} it was argued
that this full set of conditions should be taken as the definition of hybrid inflation.  

Within such models, these authors noted general features:
\begin{enumerate}
\item
The (approximate) goldstino may or may not lie in a multiplet with the
inflaton.
\item  The effective theory exhibits an approximate, continuous $R$ symmetry.
\item   Terms allowed by the discrete symmetry break the accidental continuous global symmetry
and spoil inflation, unless the inflationary scale (the square of the Goldstino
decay constant) is sufficiently small.
\item  There are further requirements on the \Kahler potential in order to obtain slow roll inflation with
adequate $e$-foldings.  This sets an {\it irreducible} minimum amount of fine tuning necessary to achieve acceptable inflation.  This tuning
grows in severity with the number of Hubble mass fields.
\item  In order that inflation ends with small {\rm c.c.}, the inflaton must couple, as noted above, to other light degrees of freedom, or must have appreciable self-couplings
in the final ground state.  The coupling to this extra field, or the self couplings, are fixed by the density perturbations $\PP_\RR$ and the inflationary scale.  In the case of extra
 fields, the resulting structure is necessarily what is called ``hybrid inflation"\cite{lindehybrid, hybrid1,hybrid2,hybrid3,linderiotto}.  The
 spectral index, quite generally, is less than one.
    \end{enumerate}

In \cite{dinepack}, it was noted that for a broad range of parameters, $n_s = 0.98$ was typical; this is widely considered a general
result of hybrid models.  Recently, considering the Planck CMB temperature data supplemented by the WMAP large-scale polarization data, the Planck collaboration has reported a value~\cite{planckinflation}:
\beq
n_s =  0.9603 \pm 0.0073.
\eeq
And indeed, the authors of the Planck papers argued that their data excludes hybrid inflation.
Within the definition outlined above, it is interesting to look more carefully at the range of allowed values of $n_s$.

In this paper, we systematically consider various Planck scale corrections to the simplest version of hybrid inflation.
We explain why (parametrically) the most important are the quartic corrections to the \Kahler potential, and certain power
law corrections to the superpotential.  The former must be suppressed by an amount of order $1/{\cal N}$, where ${\cal N}$
is the number of $e$-foldings.  The latter lead to an approximately zero c.c., supersymmetric minimum for large fields; in turn this means that the potential has a local maximum (saddle).  This gives rise to a variant of ``hilltop inflation"\cite{hilltop};
we will see that the initial conditions need not be substantially tuned in order that one obtain adequate $e$-foldings and
$n_s \approx 0.96$.  If the superpotential has coefficient scaled by a suitable power of $M_P$ and a dimensionless 
coefficient of order one, one obtains a {\it prediction} of the scale of inflation.  The scale depends on the
index $N$ of a $Z_N$ $R$ symmetry, and ranges from about $10^{11}$ GeV to $10^{15}$ GeV.  

In the next section, we review the simplest hybrid model, and recall the prediction $n_s = 0.98$.
In section \ref{hierarchy}, we classify the various Planck scale corrections to the simplest hybrid model.  In section \ref{wr}, we consider the implications of the leading superpotential corrections for inflation, explaining
why one obtains the structure of hilltop inflation.  In section \ref{numerical}, we present numerical results for these models. In section \ref{susybreaking}, we suggest that predictions might arise if inflation is connected with supersymmetry breaking. In section \ref{conclusions}, we conclude by considering possible observable consequences of this picture.

\section{Hybrid Models and $M_P$ Effects}

The simplest model of hybrid inflation contains two chiral superfields, $S$ and $\phi$, with 
superpotential
\beq
W = S(\kappa \phi^2 - \mu^2).
\label{simplestmodel} 
\eeq
If one imposes, as is usually done, a {\it continuous} $R$ symmetry under which the charges of $S$ and $\phi$ are respectively 2 and zero, this superpotential is the most general permitted by symmetries.
Classically, the theory has a moduli space, 
\beq
\vert S \vert^2 > \left| { \mu^2 \over \kappa}\right|,
\eeq
on which
\beq
V = V_0 = \vert \mu \vert^4.
\eeq
At one loop, the potential receives corrections.  In the global limit:
\beq
V = V_0( 1 + {\kappa^2 \over 8 \pi^2} \log \vert S \vert ).
\label{quantum}
\eeq
If one considers only this term, one has, for the number of $e$-foldings:
\beq
{\cal N} ={1 \over 2} {8 \pi^2 \over \kappa^2} \vert S \vert^2.
\eeq
In this simple model, the $\epsilon$ parameter is negligible, and
\beq
\eta = -{1 \over 2 {\cal N}}.
\eeq
This yields
\beq
n_s = 1- {1 \over \cal N}.
\eeq
This is the origin of the prediction that $n_s \approx 0.98$.

In this model, $\kappa$ is related to $\mu$ by the fluctuation spectrum:
\beq
\kappa = 0.17 \times \left ({\mu \over 10^{15} {\rm GeV}} \right )^2 = 7.1 \times 10^5  \times \left ({\mu \over M_P} \right )^2.
\label{kappaequation}
\eeq

\section{Hierarchy of Corrections}
\label{hierarchy}

This treatment, however, is oversimplified.  Already, in \cite{hybrid1,linderiotto}, the role of higher order terms
in the \Kahler potential was considered.  More recently, in \cite{pallisshafi}, the effects of a linear term in the potential
for $S$, arising from the constant term in the superpotential (needed to account for the small cosmological constant of the present universe)
has been considered.  In \cite{dinepack}, this particular contribution was treated as small, but a number of other effects were considered.  So it is first worthwhile to consider the various possible corrections in powers of $1/M_P$, and their
relative importance.

First, it is generally believed that theories of gravity should not exhibit continuous global symmetries; in string theories, this is a theorem.
Replacing the continuous $R$ symmetry by a discrete $Z_N$ symmetry allows corrections of the form
\beq
W_R= {\lambda \over 2(N+1) }{S^{N+1} \over M_P^{N-2}}.
\eeq

More generally, our viewpoint will be that all terms allowed in the effective action below $M_P$ should appear with order one coefficients; we will assume that smaller coefficients represent a ``fine tuning" of parameters.
We can systematically consider types of corrections, ordered in powers of $1/M_P$:
\begin{enumerate}
\item
\Kahler potential corrections:  ${\alpha \over M_P^2}(S^\dagger S)^2$, ${\beta \over M_P^4} (S^\dagger S)^3$. 
\item  Superpotential corrections:  in addition to $W_R$ (and higher powers of $S$, other fields), at some
level there must be a constant in the superpotential, $W_0$, to account for the smallness of the cosmological constant
now.
\item  Supersymmetry breaking effects.
\end{enumerate}

The term 
\beq
\delta K = {\alpha \over M_P^2}(S^\dagger S)^2
\eeq
has been noted already in \cite{linderiotto}.
In \cite{dinepack}, precise limits on $\alpha$ (of order $1/{\cal N}$, where ${\cal N}$ is the number of $e$-foldings)
were discussed.  It was noted that the quantum corrections of eqn. \eqref{quantum} only dominate over this \Kahler
potential correction for sufficiently small $S$.  In fact, as we will review shortly, for the simplest model, the quantum
corrections {\it never} dominate unless $\mu$ is quite small.

Terms of sixth order or higher in $S$ in the \Kahler potential are irrelevant.  They lead to highly suppressed contributions
to $\eta$ and $\epsilon$, for example.  We will be more quantitative about this question when we turn to models
that can reproduce the Planck value of $n_s$.

Now we turn to the various superpotential corrections.  Our definition of hybrid inflation is motivated by the
hypothesis that the scale of inflation is large compare to scales of supersymmetry breaking.  This means, in particular, that 
\beq
m_{3/2}={\vert W_0 \vert \over M_P^2} \ll H_I
\eeq
with $H_I = {\mu^2 \over M_P}$ the Hubble scale during inflation.  As a result, terms in the  potential arising from $W_0$ can be neglected during
inflation.  If, in fact, the actual value of $m_{3/2}$ is comparable to $H_I$, then this term, and terms associated with
supersymmetry breaking, would be important.  Even for $m_{3/2} = 10^2$ TeV, this corresponds to an inflationary energy 
scale well below $10^{12}$ GeV.

So finally we turn, again, to $W_R$.  The presence of this term in the superpotential gives
rise to a supersymmetric minimum of the potential at $S$ large but parametrically smaller than $M_P$.  This is unlike the case, for example, of
higher order corrections to the \Kahler potential.  As a result, this term qualitatively
alters the behavior of the system, for large but not Planck scale fields.  In \cite{dinepack} this term
was used to constrain features of inflation.  Requiring that it was not important during inflation
constrained the scale of inflation, and lead to the prediction $n_s \approx 0.98$.
To be compatible with the results from Planck, however, it is clearly necessary that inflation occur
in a region near the local maximum (as in ``Hilltop inflation"\cite{hilltop}).  We will explore this
in the next section.

\section{Hybrid Inflation and $W_R$}
\label{wr}

Including $W_R$, it is first
important that the system not flow towards the supersymmetric minimum.
Indeed, for an intermediate range of field values, there are corrections to the potential \eqref{quantum}
 of the form
\beq
\delta V_R = \lambda \mu^2 {S^N \over M_P^{N-2}}+ {\rm c.c.}
\eeq
For negative $\lambda$, this leads to a maximum, for
\beq
S^N \approx {\kappa^2 \over 8 \pi^2} {\mu^2 \over |\lambda|} M_P^{N-2}.
\eeq
To obtain suitable inflation, it is necessary that $S$ be smaller than this at the beginning.
But, given eqn. \eqref{kappaequation}, except for very large $N$, 
$S$ is smaller than the ``waterfall value",
\beq
S_w = {\mu \over \sqrt\kappa}.
\eeq

As a result of these considerations, the simplest (and rather standard) model of hybrid inflation (allowing for $W_R$) does not appear suitable.  In \cite{dinepack}, a simple
modification was suggested with two fields, $S$ and $I$, with couplings at the renormalizable level:
\beq
W = S(\kappa \phi^2 - \mu^2) +\lambda I \phi \phi^\prime + \dots
\label{modifiedhybrid}
\eeq
The theory, classically, has two flat directions, one with large $S$, one with large $I$.  As in the previous model, in order that inflation occur, the \Kahler potential
must be tuned so that at least one of the fields $S$ or $I$, has mass small compared to the Hubble constant during inflation, $H_I = {\mu^2 \over M_P}$.  To obtain
a workable model, we require that $I$ be the light field.  This amounts to requiring that in the \Kahler potential term
\beq
\delta K = \alpha \frac{S^\dagger S I^\dagger I}{M_P^2}
\eeq
$\alpha$ should be close to unity. The waterfall regime is now at smaller value of the inflaton field $I$, $I_w=\sqrt k \frac{\mu}\lambda$, and hybrid inflation can be driven by the quantum and discrete symmetry corrections.

Assuming a discrete $R$ symmetry, there are a variety of possible higher dimension terms which might appear in $W$ depending
on the transformation properties of the fields.  We will consider a term of the form
\beq
\delta W = -\gamma {SI^{N} \over M_P^{N-2}}.
\label{wri}
\eeq
Alternatively, a term proportional to $I^M$, for example,
corrects the potential for $I$ if there is a term in the \Kahler potential
\beq
\delta K = 
\frac{S I^{*M}}{M_P^{M-1}}.
\eeq
The allowed values of $M$ depend on the discrete charge assignments of the fields.  If $M$ is not too large, its effects are dramatic.

Such terms, again, lead to a supersymmetric minimum of the potential at large $I$ (with $\phi = \phi^\prime = 0$), and again give rise, for positive $\gamma$, to a {\it maximum} of the potential for $I$ at field strength generically large compared to $\mu$ but small compared to $M_P$.  

Proceeding as before, using the superpotential and \Kahler corrections in eqns. \eqref{modifiedhybrid}--\eqref{wri}, we can compute the number of $e$-foldings and the slow roll parameters (and hence $n_s$).  
The potential for $I$ is now, approximately,
\beq\label{potN}
V(I) = \mu^4 \left(1 + {\kappa^2 \over 16 \pi^2} \log(I^\dagger I) - (\alpha-1)\frac{I^\dag I}{M_P^2}\right)-{\gamma \mu^2 M_P^2} \left ( {I \over M_P} \right )^N + {\rm c.c.}.
\eeq

The fluctuation spectrum relates $\kappa$ and $\mu$, as before.  For a given value of $\mu$, the initial value of the field at $\NN=60$ $e$-foldings is fixed.
So, then, is $n_s$.  

To get a rough sense of scalings, we can suppose that $I$ starts very near the maximum of the potential, and that $\eta= -0.02$ (in order to achieve $n_s=0.96$).
Because $V^\prime \sim 0$ at the hilltop, we will simply use the formula for normal hybrid inflation in our estimate;
shortly we will check the accuracy of this numerically, and see that this leads to an order one error. Then one finds that
\beq
 {\mu \over M_P}  = 
\left (\left({.02 \over N}\right)^N\cdot  (6.4 \times 10^9)^{2-N} (N\gamma)^{-2} \right )^{1 \over 4N -12}.
\eeq
%
For particular values of $N$, we can compute $\mu$  and $\kappa$: taking $\gamma \approx 1$ and $N=4$, this gives $\mu \approx 10^{11}$ GeV and $\kappa \approx 10^{-10}$. For $N=5$, one obtains $\mu \approx 10^{13}$ GeV, and $\kappa \approx 10^{-5}$. The scale $\mu$ grows slowly with $N$, reaching $10^{14}$ GeV  at $N=7$ and $10^{15}$ GeV for $N=12$. In general, these results scale with $\gamma$ as:
\beq
\gamma^{-{1 \over 2(N-3)}}.\eeq
We discuss numerical studies of this problem in the next section.  But the lesson here is that, for fixed values of $\gamma$,  and for a given $N$, the scale of inflation, $\mu$, is fixed to a narrow range.

\section{Numerical Studies of Small Field Inflation}
\label{numerical}

Denoting the real part of the field $I$ by $\sigma$, the potential in eqn. \eqref{potN} becomes
\beq\label{potNsigma}
V(\sigma) = \mu^4 \left(1 + {\kappa^2 \over 16 \pi^2} \log(\sigma^2) - (\alpha-1)\frac{\sigma^2}{M_P^2}\right)-{\gamma \mu^2 M_P^2} \left ( {\sigma \over M_P} \right )^N,
\eeq
where we have included in $\gamma$ the numerical factor $2^{N/2-1}$ coming from the field redefinition.
It will be handy to denote the hilltop position by $\sigma_h$, and to investigate how close $\sigma$ has to be to $\sigma_h$ in order to successfully have $\NN=\ $50--60 $e$-foldings of inflation.

For a given $N$, the parameters of the two field model are readily enumerated:
 $\mu$, $\kappa$, $\alpha$, and $\gamma$. 
Given knowledge of these, we can compute the observable predictions of the inflationary model, to be compared with the Planck collaboration results \cite{planckinflation}:
\begin{enumerate}
\item  The number of $e$-foldings $\NN$. To solve the horizon and flatness problems, it must be $\NN\geq 50$. In our numerical treatment, we will assume the range of $\NN=\ $50--60 $e$-foldings.
\item  The slow roll parameters $\eta, \epsilon$, which result in the spectral index $n_s=1-6\epsilon+2\eta$. The measured value by the Planck collaboration is $n_s =  0.9603 \pm 0.0073.$
\item  The density perturbation spectrum $\PP_\RR$, whose amplitude is a function of $V^{3/2}/V'$. Planck measurements translate to $V^{3/2}/V'=(5.10 \pm 0.07) \times 10^{-4}M_P^3$.
\end{enumerate}

We can, in principle, compute the tensor to scalar ratio $r$, but in all such models this will be unobservably small. In general, as said in the previous section, $(1-\alpha)$ quantifies the \Kahler correction independent from the discrete symmetry, and is already required to be small, while the dependence on $\gamma$ is weak. In the following we will set $\alpha\sim1$, $\gamma=1$.

\begin{figure}[tbp]
\begin{center}
\includegraphics[height=10cm]{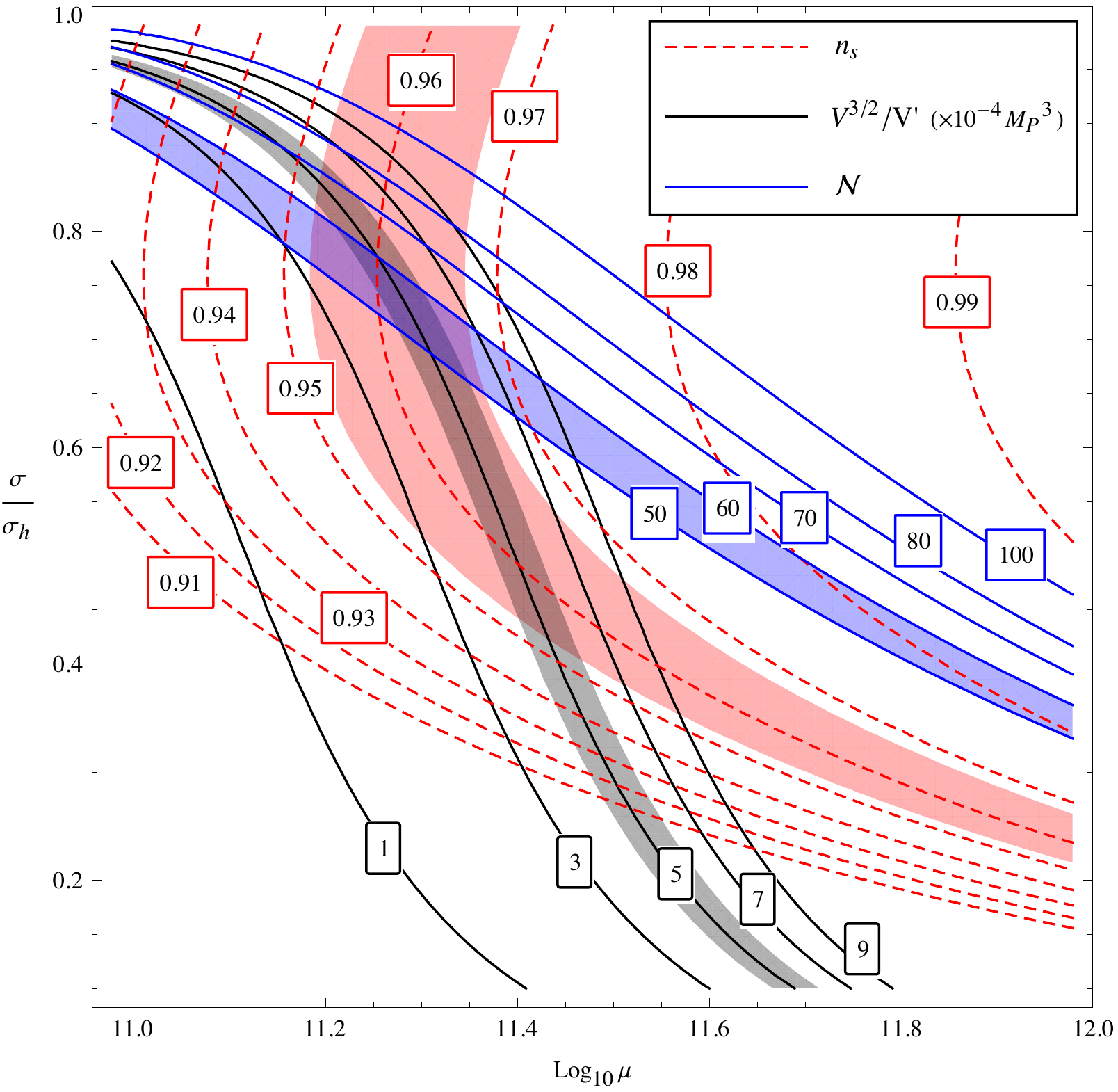}
\caption{Contours for the spectral index $n_s$ (dashed red), the density perturbation $V^{3/2}/V'$ (solid black) and the number of $e$-foldings $\NN$ (solid blue), for $N=4$. The coupling $\kappa$ is kept fixed at its best fit value of $\kappa=2.1\times 10^{-9}$. The shaded zones indicate the 1-sigma regions allowed by the Planck results for $n_s$ and $V^{3/2}/V'$, and the range of 50--60 $e$-foldings. The $\chi^2$ is minimized where the bands intersect each other. For each value of $\kappa$ a specific range of $\mu$ and $\sigma$ is allowed. As $\kappa$ varies, each variable changes independently and the allowed region moves and shrinks, until the three bands do not intersect.
}
\label{N4allcontours}
\end{center}
\end{figure}

Given the potential \eqref{potNsigma}, the expression for the number of $e$-foldings $\NN$ involves an integral that can be computed numerically. With a $\chi^2$ analysis, for each given $N$, we set the three remaining parameters $\mu,\sigma,\kappa$ by fitting the experimental values of $\NN, n_s, V^{3/2}/V'$. For example, in Fig. \ref{N4allcontours}, where we set $N=4$ and $\kappa$ to its best-fit  value, we show how the allowed ranges for each experimental quantity intersect at specific values of $\mu$ and $\sigma$.

In Table \ref{numtable}, we give the best-fit values of $\mu, \sigma, \kappa$, with the corresponding uncertainties for $N$ from 4 to 12. There is no fine-tuning associated with the inflaton being close to the hilltop value, as the allowed values for $\sigma/\sigma_h$ are in the range $0.6$--$0.8$. For small $N$, the coupling $\kappa$ is tuned to be small. In the last column, we show how close to unity $\alpha$ has to be for the \Kahler correction not to overcome the discrete symmetry correction. As $|\alpha-1|$  is already  fine-tuned to be of order $\frac1\NN$ in order not to spoil inflation, we conclude that there is another mild tuning which operates to keep $\alpha$ close to 1.

\begin{table}[tdp]
\begin{center}
$
\begin{array}{|c||c|c|c|c|c|}\hline
N  & \mu\ (\text{GeV}) & \sigma\ (\text{GeV})& \sigma_h\ (\text{GeV}) & \kappa &|\alpha-1|_{max}   \\\hline\hline
4   &(2\pm0.5)\times10^{11}& (4.8\pm2.3)\times10^9 &6.7\times10^9& (2.1\pm1.2) \times 10^{-9} &0.004 \\\hline
5   &(2.5\pm0.5)\times10^{13}& (8\pm2)\times10^{13} &11\times10^{13}& (3.5\pm1.5) \times 10^{-5} &0.006 \\\hline
6   &(1.25\pm0.2)\times10^{14}& (2.1\pm0.4)\times10^{15} &2.8\times10^{15}& (9\pm1) \times 10^{-4}  &0.007 \\\hline
7   &(2.9\pm0.5)\times10^{14}& (1.1\pm0.16)\times10^{16} &1.42\times10^{16}& (4.7\pm0.7) \times 10^{-3} & 0.008 (-);0.02 (+) \\\hline
8   &(4.8\pm0.7)\times10^{14}& (2.9\pm0.3)\times10^{16} &3.8\times10^{16}& (1.3\pm0.1) \times 10^{-2}& 0.01(-) ;0.016(+)  \\\hline
12   &(1.2\pm0.15)\times10^{15}& (1.74\pm0.12)\times10^{17} &2.14\times10^{17}& (7.4\pm0.6) \times 10^{-2} &0.013 (-);0.008(+) \\\hline
\end{array}
$
\end{center}
\caption{Numerical results: central values and 1$\sigma$ allowed ranges for the parameters, for different choices of $N$. The central column lists the hilltop value for the central value of the parameters. The last column shows how close to 1 the quartic Kh\"aler correction $\alpha$ is forced to be (at the 95\%CL); for some $N$, there is a weak dependence on the sign of ($\alpha-1$); these values should be compared to the irreducible tuning of order $\frac1\NN\sim 0.016$--$0.020$.
}\label{numtable}
\end{table}

For $N=12$, the initial value of the field is  $\sigma=1.7 \times 10^{17}$ GeV, just a factor of 10 below $M_P$. For larger $N$, it is not possible to accomodate $n_s=0.96$ within the framework of small field inflation. Even for this large value of the field, the tensor-to-scalar ratio is predicted to be small:
\beq
r=0.12 \frac{V}{(1.94\times 10^{16}\text{ GeV})^4}\leq2\times 10^{-6}
\eeq

\section{Incorporating Supersymmetry Breaking}
\label{susybreaking}

The picture of small field inflation we have developed up to now assumes that the scale of inflation is large compared to the scale of supersymmetry breaking, i.e. that $H_I \gg m_{3/2}$.  This is the origin of the requirement that the superpotential should vanish and supersymmetry be unbroken, to a good approximation, at the end of inflation.  But one might consider the possibility that $H_I \sim m_{3/2}$.  A higher scale of $m_{3/2}$ is suggested
by the observed Higgs mass and supersymmetry exclusions.  In addition, for small values of $N$, we have obtained small values of
$H_{0}$.  So it is interesting to consider the possibility that the the scale of inflation is comparable to $m_{3/2}$.

For example we can modify the models we have studied, to give them an O'Raifeartaigh like structure, adding to the superpotential of
eqn. \eqref{simplestmodel}
a coupling
\beq
m \phi \Phi.
\eeq
Provided
\beq
\vert m^2 \vert > \kappa \mu^2
\eeq
supersymmetry is broken, in a state with $\Phi = 0$.
It is interesting that in this case, inflation ends without ever passing into a ``waterfall" regime.  As we have stressed, the so-called waterfall is indeed not the distinguishing feature of hybrid inflation.

A different approach has been pursued in \cite{Takahashi:2013cxa}.  Again, it is assumed that the scale of inflation is not too much different than the scale of supersymmetry breaking.  One writes a theory of a single field, $\phi$, and does not require an unbroken $R$ symmetry at the end of inflation.  Instead, one assumes that the negative contribution to the cosmological constant arising from the vev of the superpotential is cancelled by some supersymmetry breaking dynamics.  To constrain the form of the superpotential, one still assumes a discrete $R$ symmetry.  It is necessary, as in hybrid inflation,
to tune the \Kahler potential so that the $\vert \phi \vert^4$ term is small.
The superpotential takes the form:
\beq
W(\phi) = v^2 \phi - {g \over n+1} \phi^{n+1},
\eeq
while the quartic term in the \Kahler potential must be quite small.
The resulting model is of the hilltop type.  The potential exhibits a local maximum at the origin, and the initial value of the field must lie quite close to the maximum  (compared to the distance of the origin from the minimum).  Inflation occurs in a region very close to the origin in field space (defined by an unbroken $R$ symmetry).  The field then settles into a minimum with small cosmological constant and broken supersymmetry and $R$ symmetry.  The model can produce the requisite number of $e$-foldings and fluctuation spectrum, without introducing an extremely small number analogous to $\kappa$ of eqn. \eqref{simplestmodel}.  However, it predicts too {\it small} a value of $n_s$,
\beq
n_s = 0.94.
\eeq
To obtain a spectral index consistent with Planck, it is necessary to introduce a small and well-tuned constant in the superpotential, which
the authors denote $c$, and is of order $10^{-19}$ (in Planck units).  There are other issues, such as a possible gravitino problem and overproduction of dark matter, but these can readily be solved by introducing additional matter coupled to the inflaton.

Both approaches, then, seem viable, and have the potential to relate supersymmetry breaking dynamics to inflationary dynamics.  Each requires certain tunings.

\section{Conclusions:  Predictions and Observable Consequences for Low Energy Physics}
\label{conclusions}

The results from Planck pose challenges for models of small field inflation.  It has been said that they rule out ``hybrid inflation."  Here, following \cite{dinepack}, we have carefully defined models of hybrid inflation as models in which inflation occurs on a pseudomoduli space, with supersymmetry and an $R$ symmetry approximately restored at the end of inflation.  We have assumed a {\it discrete} $R$ symmetry, and have considered the importance of corrections to the superpotential and \Kahler potential.  For initial values of the field far from the local maximum of the potential, one predicts a spectral index inconsistent with Planck.  To obtain $n_s = 0.96$, it is necessary
that the field start near the local maximum, though this condition is not severely tuned.  For $Z_N$ symmetry with $N=4$, the scale of inflation is rather low, and we considered the possibility that $H_I \approx m_{3/2}$.  In this case, the dynamics of inflation might be closely tied to the scale of supersymmetry breaking, and there is some chance that aspects of the physics of inflation could be studied in accelerator experiments.  

We have noted that, in this case, the assumption of an unbroken $R$ symmetry and unbroken supersymmetry at the end of inflation might be relaxed, and compared the hybrid models with those of \cite{Takahashi:2013cxa}.  Each of these models can reproduce the data, and involves very small parameters and tunings.  The fact that many models with such features can reproduce the basic data of inflation raises, as always, the question of whether there is any way they might be testable or falsifiable.  We would argue that the best hope is connecting inflation with the dynamics responsible for supersymmetry breaking.  It will be particularly interesting to explore dynamical supersymmetry breaking (and generation of scales) in this framework.

\noindent
{\bf Acknowledgements:}  We thank Andrei Linde and Fuminobu Takahashi for discussions.
This work supported in part by the U.S.
Department of Energy.

%

\end{document}